\documentstyle[aps,prl,epsfig]{revtex}
\tightenlines
\begin{document}
\twocolumn[\hsize\textwidth\columnwidth\hsize\csname@twocolumnfalse\endcsname
\draft
\title{Comment on "Peculiar Scaling of Self-Avoiding Walk Contacts"}
\maketitle
%\pacs{05.70.Jk,36.20.Ey,64.60.Ak,64.60.Kw}
\vfill
\twocolumn
\vskip.5pc]
\narrowtext

Baiesi, Orlandini and Stella have recently shown [1] that the set of all
non-bonded nearest neighbour (NN) contacts between the two halves of a 
self-avoiding walk is a multifractal characterized by a universal 
exponent, ${\tau}$. In the presence of NN interactions, and in two 
dimensions, they have shown that $\tau \sim 1.93$ 
above the $\theta $ temperature, $T_{\theta}$, wheras $\tau =1$
at and below $T_{\theta}$. They have shown further that the exponent
$\tau$ could be superuniversal with a value, ${\tau} = 2$, for contacts 
between a self-avoiding walk and a confining wall. The purpose of this 
comment is to demonstrate that such contact-scaling is broken for the recently 
proposed Interacting Growth Walk (IGW) [2].

IGW starts from an arbitrarily chosen site, ${\bf r}_{0}$, of  a  regular  
$d$-dimensional  lattice  of coordination  number  $z$, whose sites are 
initially 'unoccupied' (by monomers). The first step is made in  
one  of the  $z$  available directions, by choosing an 'unoccupied' NN of
${\bf r}_{0}$, say ${\bf r}_{1}$, at  random with equal probability.  
Let  $\{  {\bf  r}_{j}^{m} \mid  m=1,2,...,z_{j}  \}$  be the 'unoccupied' 
NNs available for the $j^{th}$ step of the walk. If $z_{j}=0$,  the  walk  
can  not grow  further  because  it  is  geometrically  'trapped'.  It is,
therefore, discarded and a  fresh  walk  is  started  from  ${\bf
r}_{0}$.  If $z_{j} \neq 0$, the walk proceeds by choosing one of
the available sites, say ${\bf r}_{j}^{m}$, with a probability proportional to
$exp [-\beta n_{NN}^{m}(j) \epsilon _{0}]$, where 
$n_{NN}^{m}(j)$ is the  number  of  non-bonded NN  sites  of  ${\bf
r}_{j}^{m}$ and $\beta = 1/k_{B}T$, $k_{B}$ is the Boltzmann constant.
At 'infinite'   temperature   ($\beta   =0$), the  walk  generated  will  be  the  
same as the Kinetic Growth Walk (KGW), reference given in [2]. However, at
finite temperatures, the walk will prefer to step into a site
with  more non-bonded NN contacts when the energy
per contact, $\epsilon_{0}$, is negative. We may set $\epsilon _{0}= -1$ 
without loss of generality. 

At any given temperature, we make a large number of attempts [3] to generate 
$N$-step IGWs, identify the non-bonded NN contacts between the two halves of the 
walks, locate their centres of mass and compute the ensemble average of the $q$-th 
power of the average radius of gyration, $<R_{c}^{q}(N)> \sim N^{\sigma _{q}}$, $(q>0)$. 
From the observed power-law dependence of the moments, we obtained 
$\sigma_{q}$ as a function of $q$ for IGWs on a square lattice, and have 
plotted them in the upper inset of Fig.1 for $\beta = 0, 1/2, 1, 2$ and $300$.
The fact that $\sigma _{q}/q$ depends on $q$ for all these temperatures is related to 
the specific nature of the corresponding distributions, $P_{\beta}(R_{c},N)$, 
which we have shown in the lower inset of Fig.1 for $N=300$. It is clear that the distribution has 
a power-law form, $P_{\beta = 0}(R_{c},N=300) \sim R_{c}^{-\tau}$, only 
at infinite temperature. By demanding that the $q$-th moment, obtained from the scaling form, 
$P_{\beta = 0}(R_{c},N) \sim R_{c}^{-\tau}f( R_{c}/N^{\nu})$ where the end-to-end
distance exponent $\nu = 3/4$ for the KGW, be consistent with the monte carlo data
in Fig.1, we get the contact-exponent, $\tau \sim 1.90 \pm 0.02$. That this value of
$\tau$ is consistent with the assumed scaling form is further confirmed in  
Fig.1, where we have shown that the distributions corresponding to $N = 200$ and 
$N = 300$ could be collapsed reasonably well. Thus, the IGW at infinite temperature 
(or equivalently, the KGW) has the same contact-scaling behaviour as a 
self-avoiding walk However, at finite temperatures, it has none.

%%%%%%%%%%%%%%%%%%%%%%%%%%%%%%%%%%%%%%%%%
%                       FIG.1                                                                                               %
%%%%%%%%%%%%%%%%%%%%%%%%%%%%%%%%%%%%%%%%%
\begin{figure}
\includegraphics[width=3.3in,height=2.4in]{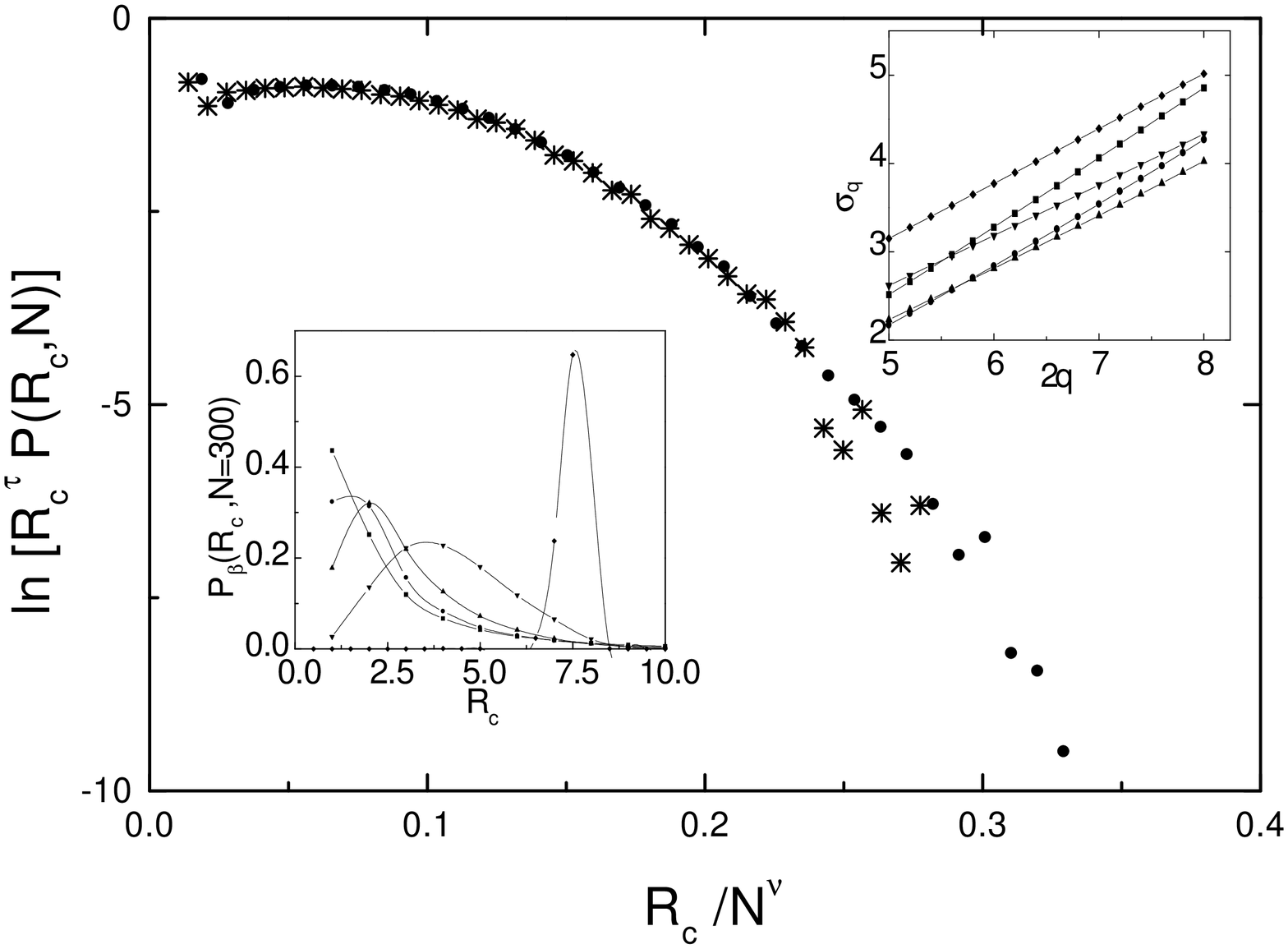}
\caption{Data collapse for $N=200$ (filled circle) and $N=300$ (star) at infinite temperature ($\beta =0$) 
with the contact-exponent $\tau =1.9$. Upper Inset: $\sigma _{q}$ as a function of $q$
for $\beta = 0, 0.5, 1, 2$ and $300$. Lower Inset:  The distributions, $P_{\beta}(R_{c},N=300)$,
for $\beta = 0, 0.5, 1, 2$ and $300$ from left to right.}
\end{figure}
%%%%%%%%%%%%%%%%%%%%%%%%%%%%%%%%%%%%%%%%%

\medskip
\medskip
\noindent S.L. Narasimhan$^{1*}$,  P.S.R. Krishna$^{1*}$, K.P.N.Murthy$^{2}$
and M. Ramanadham$^{1}$ \\
$^1$Solid State Physics Division,
Bhabha Atomic Research Centre, Mumbai - 400 085, India.
\\$^2$Materials  Science  Division, Indira Gandhi Centre for Atomic Research,
Kalpakkam - 603 102, India.

\noindent $^*$glass@apsara.barc.ernet.in
%\pacs{05.70.Jk,36.20.Ey,64.60.Ak,64.60.Kw}


\begin{references}
\bibitem{1} M. Baiesi, E. Orlandini and A. L. Stella, Phys. Rev. Lett. {\bf 87},070602 (2001).
\bibitem{2} S. L. Narasimhan, P. S. R. Krishna, K. P. N. Murthy and M. Ramanadham, preprint {\it cond-mat}/0108097,
and references therein; submitted to Phys. Rev. E. 
\bibitem{3} Typically $\sim 200$ million attempts are made to generate walks upto a maximum
of $N = 500$ on a square lattice; sample  attrition is the most severe problem for 
$\beta=0$ and it becomes less and less severe as the value of  $\beta$  increases.
\end{references}
\end{document}